\documentstyle[11pt,IAUS215,twoside,epsf]{article}
\def\chem#1#2{$\rm{}^{#2}\kern-0.8pt#1$}
\def\reac#1#2#3#4#5#6{$\rm\,{}^{#2}\kern-0.8pt{#1}\,({#3}\,,{#4})\,
{}^{#6}\kern-0.8pt{#5}\,$}
\def\gsimeq{\,\,\raise0.14em\hbox{$>$}\kern-0.76em\lower0.28em\hbox  
{$\sim$}\,\,}  
\def\lsimeq{\,\,\raise0.14em\hbox{$<$}\kern-0.76em\lower0.28em\hbox  
{$\sim$}\,\,}  
\def\beqy{\begin{eqnarray}}
\def\eeqy{\end{eqnarray}}
\def\bmlet{\begin{mathletters}}
\def\emlet{\end{mathletters}}

\def\edcomment#1{\iffalse\marginpar{\raggedright\sl#1\/}\else\relax\fi}
\reversemarginpar

\begin{document}
\title{The effects of rotation on Wolf--Rayet stars and on the production of
primary nitrogen by intermediate mass stars}
 \author{Georges Meynet$^1$ and Max Pettini$^2$}
\affil{1) Geneva Observatory, CH--1290 Sauverny, Switzerland}
\affil{2) Institute of Astronomy, Madingley Road, Cambridge, CB3 0HA, UK}
 
\begin{abstract}
We use the rotating stellar models described in the paper by A. Maeder \& G. Meynet
in this volume to consider the effects of rotation on the evolution 
of the most massive stars into and during the Wolf--Rayet phase, 
and on the post-Main Sequence evolution of intermediate mass
stars. The two main results of this discussion are the following. 
First, we show that rotating models are able to account for the observed 
properties of the Wolf--Rayet stellar populations at 
solar metallicity. Second, at low metallicities, the inclusion
of stellar rotation in the calculation of chemical yields can 
lead to a longer time delay between the release of oxygen and nitrogen
into the interstellar medium following an episode of star formation,
since stars of lower masses (compared to non-rotating models)
can synthesize primary N. Qualitatively, such an effect may be
required to explain the relative abundances of N and O in extragalactic 
metal--poor environments, particularly at high redshifts.

\end{abstract}

\section{The Effects of rotation on the evolution into the Wolf--Rayet phase}

As is recalled in the paper by P. Eenens in this volume, Wolf--Rayet (WR) 
stars are the bare cores of initially massive stars,
whose H--rich envelope has been removed by strong stellar winds or 
through Roche lobe overflow in a close binary system.
Here we consider the following question:
what are the effects of rotation on the evolution of massive single 
stars into the Wolf--Rayet phase? This subject has been discussed by Maeder (1987),
Fliegner and Langer (1995), Maeder \& Meynet (2000) and Meynet (2000). 
We shall briefly summarise the main results
presented in those papers and assess their importance 
in the framework of a new grid of massive star models at solar 
metallicity (Meynet \& Maeder in preparation).

As a preamble, let us reconsider the criteria which have been chosen 
to decide when a stellar model enters into the WR phase. 
Ideally, of course, the physics of the models should decide 
when the star is a WR star. However our poor knowledge of the physics 
involved, as well as the complexity of models
coupling the stellar interiors to the winds, are such that this 
approach is not yet possible.
Instead, it is necessary to adopt some empirical criteria
for deciding when a star enters the WR phase. 
In this work the criteria are the following: 
the star is considered to be a WR star when its temperature is
$\log T_{\rm eff} > 4.0$
and the mass fraction of hydrogen at the surface is $X_S < 0.4$. 
Reasonable changes to these values (for instance adopting $X_S < 0.3$ 
instead of 0.4) do not affect the results significantly.

\begin{figure} 
\plotone{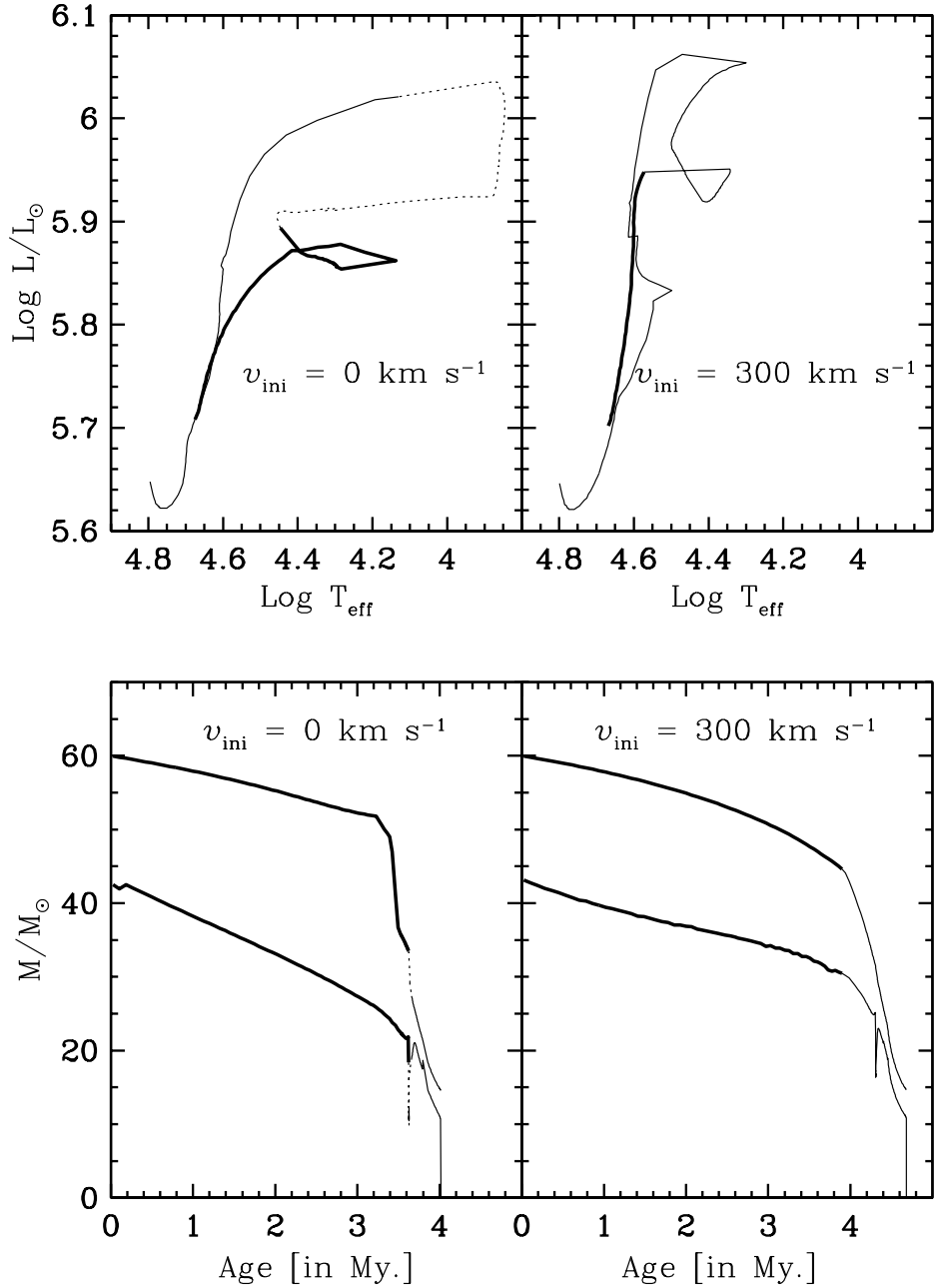}
\caption{The two upper panels show the evolutionary tracks of a non--rotating and a rotating 60 M$_\odot$ stellar model at solar metallicity.
The bold part of the track corresponds to the O--type star Main--Sequence phase; the dotted part show the intermediate phase,
when present, between the O--type star phase and the WR phase; the thin continuous line is the track during
the WR phase. The two lower panels show the evolution as a function of time of the total mass and of the masses of the
convective cores during the H-- and the He--burning phases.}
\end{figure}

Figure 1 shows the evolutionary tracks and the evolution of the structure in a non--rotating and a rotating 60 M$_\odot$ model. 
The rotating model has a time--averaged equatorial velocity on the Main Sequence (MS) of about 190 km s$^{-1}$, 
not far from the mean equatorial velocity observed for O--type stars. Inspection of 
Fig. 1, shows the following differences:
\vskip 2mm
\begin{itemize}
\item The non--rotating 60 M$_\odot$ star goes through an intermediate Luminous Blue Variable (LBV) phase
before becoming a WR star, while the rotating model skips the LBV phase and goes directly from the O--type star 
phase into the WR phase. One notes also that during the O--type star phase, the rotating track is bluer than its non--rotating counterpart. 
This is due to the diffusion of helium into the outer radiative zone (see Heger \& Langer 2000;
Meynet \& Maeder 2000). On the other hand, the luminosity during the WC phase is not different between the rotating and the non--rotating model. 
Since these stars follow a mass--luminosity relation (Maeder 1987; Schaerer \& Maeder 1992),
this implies that the masses of these two stars at this stage are about equal. 
At first sight this may be surprising. One expects the rotating star to lose more mass than its non--rotating counterpart. 
However, in this particular case there are some compensating effects. Indeed,  
the non--rotating model enters into the WR phase later, but becomes redder during the MS phase and then becomes a LBV star.
During these two last phases, the star undergoes strong mass loss (see the left panels of Fig. 1). 
The rotating model, on the other hand, enters 
the WR phase at an earlier stage, reaching at the end an identical mass to the non--rotating model.
In general, however, the final masses obtained from the rotating models are smaller than those obtained from the non--rotating ones, 
leading to lower luminosities during the WC phase.

\item Focusing on the lower panels of Fig. 1, it can be realised that the non--rotating star enters 
in the WR phase with a mass of about 27 M$_\odot$. 
Nearly the whole H--rich envelope has been removed by stellar winds during the previous phases 
(more precisely at the end of the MS and during the LBV phase). 
In this case, the main mechanism responsible for making the star a WR star is 
mass loss by stellar winds. For the rotating model, on the other hand,
the entrance into the WR phase occurs at an earlier stage 
(although not at a smaller age!), while the star is still burning hydrogen in its core. 
The mass of the star at this stage is about 45 M$_\odot$
and an important H--rich envelope is still surrounding the convective core.
The main effect responsible for making the star a WR star in this second case is
rotationally induced mixing (Maeder 1987; Fliegner \& Langer 1995, Meynet 2000).

\end{itemize}
\vskip 2mm

\begin{figure} 
\plottwo{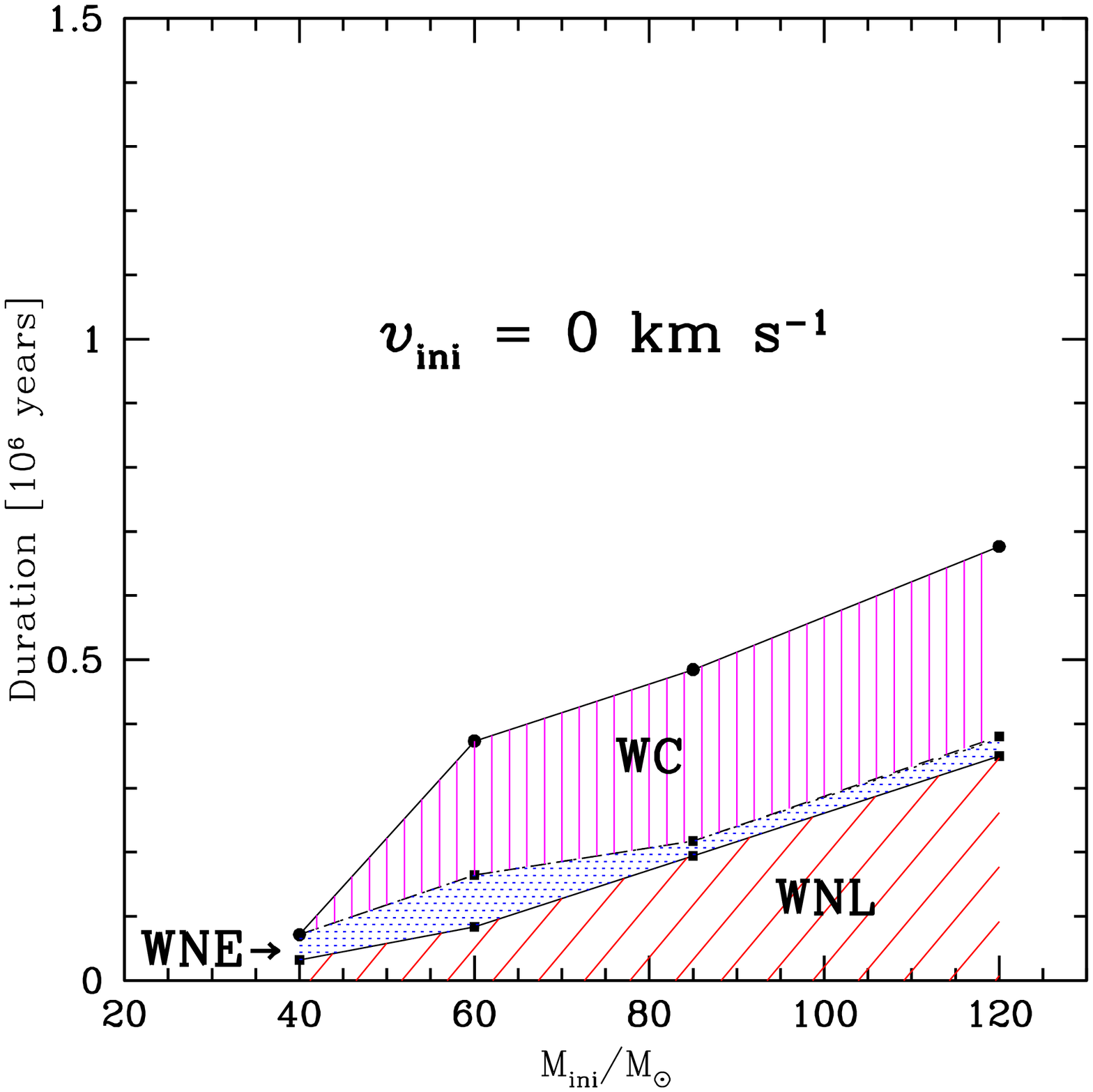}{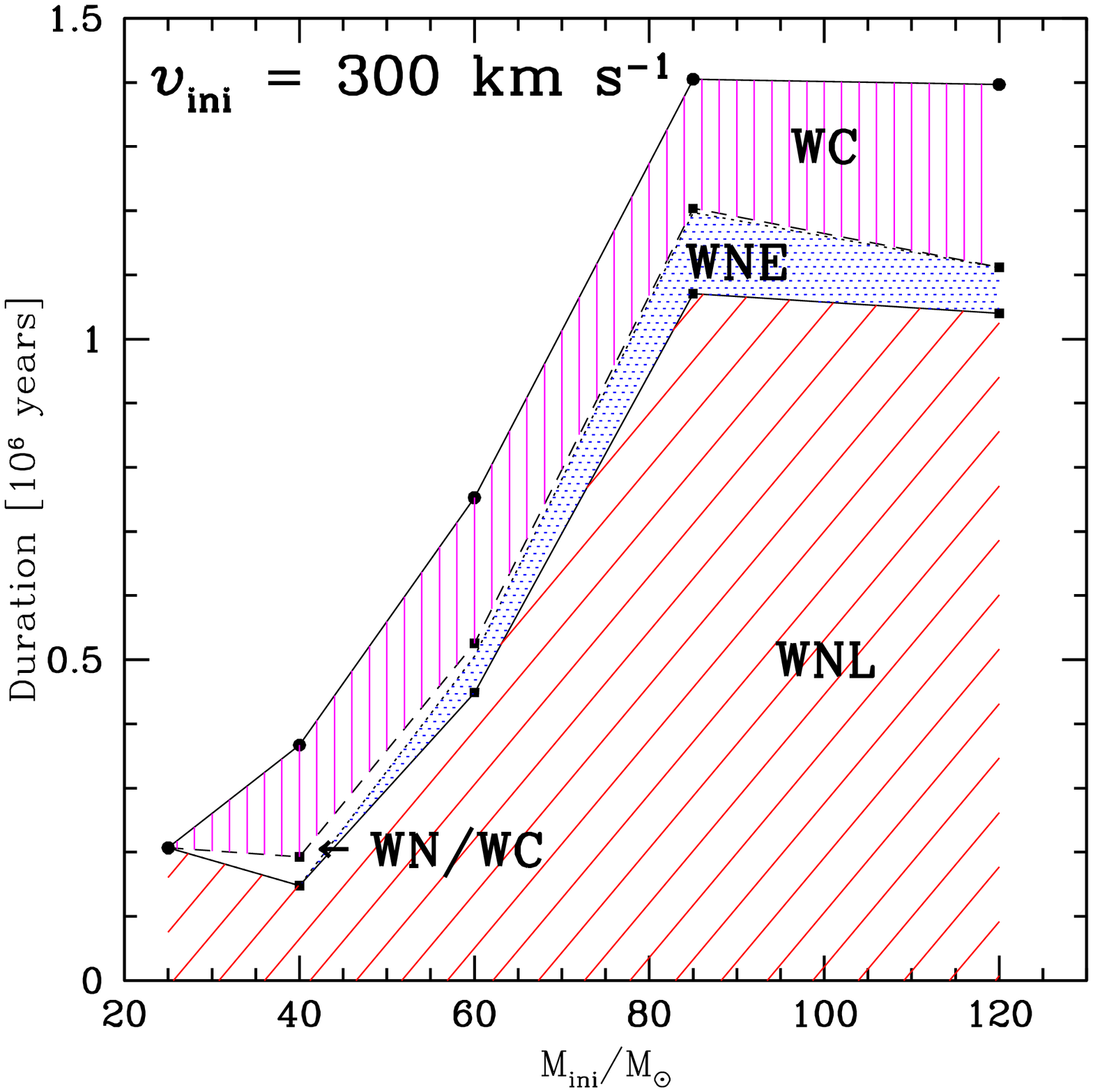}
\caption{Durations of the WR phases and of the WR sub-phases for stars of various initial masses. The left panel
shows the case for non--rotating models while the right panel corresponds to the case of rotating models.}
\end{figure}

These differences between the behaviours of the rotating and the 
non--rotating models have important consequences for the duration of the WR phase as a whole
and for the lifetimes spent in the different WR subtypes, as can be appreciated from Fig. 2.
\vskip 2mm
\begin{itemize}
\item First, the WR lifetimes are enhanced. The typical enhancement 
for the 60 M$_\odot$ model amounts to nearly a factor of 2.

\item Second, the WNL phase (the phase during which CNO processed material
appears at the surface together with H) is considerably lengthened.
This is a natural consequence of the fact that, when the rotating stellar model 
enters the WR phase, it still has an important H--rich envelope. 
The WNL phase duration increases with the initial velocity.
When the initial velocity passes from 300 to 500 km s$^{-1}$ 
(or when the time--averaged equatorial velocity during the O--type phase goes 
from 190 to 330 km s$^{-1}$), the WNL phase duration is increased by a factor of about 1.7.
The durations of the WNE and WC phases are, on the other hand, affected much less 
(the WNE phase corresponds to the time when one sees at the stellar surface CNO processed material 
without hydrogen; during the WC phase, He--burning products such as carbon and oxygen appear at the surface).

\item Thirdly, in the rotating stellar model a new phase of modest,
but of non--negligible duration, appears: the so--called transitional WN/WC phase. 
This phase is characterized by the simultaneous presence at the surface of both
H-- and He--burning products (typically N and C or Ne). The reason for this
are the shallower chemical gradients which build up inside the rotating models.
These shallower gradients inside the stars also produce a smoother evolution of the surface abundances 
as a function of time (or as a function of the remaining mass, see Fig. 7 in Maeder \& Meynet 2000). 
For a transitional WN/WC phase to occur, it is necessary to have---for a sufficiently long period---both 
a He--burning core and a CNO--enriched envelope.
In the highest mass stars, mass loss removes too rapidly the CNO--enriched envelope 
to allow a transitional WN/WC phase to appear. 
In the low mass range, the time spent in the WR phase is 
too short and the H--rich envelope too
extended to allow He--burning products to diffuse up to the surface. 
Consequently, the transitional WN/WC phase only appears in the mass range between
30 and 60 M$_\odot$. 

\item Finally, the minimum mass for a star to become a WR star (through the single star channel) 
is lowered by rotation. In the present case, the minimum mass is reduced from about 
37 M$_\odot$ for $v_{\rm ini}=0$ km s$^{-1}$
to about 22 M$_\odot$ for $v_{\rm ini}=300$ km s$^{-1}$.

\end{itemize}

From the above considerations, one can conclude that, compared
to non-rotating models, stellar models including rotation will
predict higher values of: (a) the numbers of WR stars relative to O--type stars, 
(b) the relative number of WN to WC stars, and (c) the relative number of transition WN/WC stars to WR stars. 
Are the observed values of these ratios in better agreement 
with the predictions of the rotating or non--rotating models? 
Clearly rotating models are in much better agreement.

However, before discussing further this conclusion, 
it is worthwhile recalling
some of the difficulties faced by non--rotating stellar models. 
It was shown by Maeder \& Meynet (1994) 
that the relative numbers of WR and O-type stars observed in regions of 
different metallicities could be well
reproduced by the stellar models, provided the mass loss rates were enhanced 
by about a factor of two during the MS and WNL phases. 
This is true not only for regions of constant star formation rate, 
but also for starburst regions
(Schaerer et al. 2000). 
This enhancement appeared to be a reasonable assumption
at the time of that study, given the 
uncertainties in the mass loss rates determinations available then.
However, this solution is no long applicable today. 
Indeed it is now well established that the mass loss rates 
during the WR phase should be {\it reduced} by a factor of 2--3 
owing to clumping effects in the wind (Nugis and Lamers 2001; Hamann et al. 1999). 
Thus we are left with the following problem: how can one account for the observed WR populations 
without increasing artificially the mass loss rates?  
The solution presumably lies in some physical process, not accounted for in the models of 1994, 
which would act as an enhancement of the mass loss rates as far as WR star formation is concerned.

Another difficulty faced by standard non--rotating models is the observed fraction of 
WR stars in the transition WN/WC phase. 
Although modest (a few percent), this fraction is much greater 
than predicted by non--rotating stellar models (with or without enhanced mass loss rates). 
This last discrepancy clearly points towards the need for an extra--mixing mechanism, 
as proposed by Langer (1991).

From our discussion above it is evident that rotation may help in solving these problems. 
Like enhanced mass loss rates, rotation lengthens the WR phase and decreases the minimum 
mass for a single star to go through a WR phase.
 Adopting the O and WR lifetimes obtained from the rotating models, the WR/O, WN/WC and (WN/WC)/WR number
ratios at solar metallicities are well reproduced (Meynet \& Maeder in preparation), 
while predictions for these same quantities obtained from non--rotating models 
are in clear disagreement with the observed values. Thus, one mechanism---rotational mixing---helps 
in solving both the problem of the WR/O number ratio and that of the transition WN/WC phase. 
It is somewhat encouraging to see that inclusion in stellar models of one observed characteristic 
of massive stars, namely their fast rotation, improves significantly the match of the models
to reality.

We conclude this section on WR stars with a few additional  remarks.
\vskip 2mm
\begin{itemize}
\item For the moderate velocities considered here ($v_{\rm ini}$ = 300 km s$^{-1}$
on the ZAMS), the effects of wind anisotropies (Maeder \& Desjacques 2001)---which
are accounted for in the present models---play a modest role.

\item Rotation may help in producing WR stars from single stars in metal poor environments,
where the mass loss rates are expected to be much lower than at solar metallicity.
This can result for at least two reasons. 
First, rotational diffusion can deplete the surface hydrogen abundance, 
making the star enter the WR phase well before mass loss has uncovered the core of the star. 
Second, since mass loss removes less angular momentum from the surface, 
the star can reach the break--up limit more easily. 
In this case, important mass loss rates ensue and this may help the star enter into the WR phase. 

\item As discussed by P. Eenens, very little is known about the rotation of WR stars.
The present models predict that the equatorial velocity is between 40 and 70 km s$^{-1}$ during the WR phase. This
corresponds to rotational periods of about 70 days during the WN phase and of a few days during the WC phase.
A more complete discussion of this point will be found in the paper by Meynet \& Maeder (in preparation).
\end{itemize}

\section{Primary nitrogen production by rotating metal--poor intermediate--mass stars}

Primary nitrogen is produced by the transformation, 
through the CNO cycle in H--burning zones, 
of carbon and oxygen synthesised by the star itself in He--burning regions. 
Nitrogen also has a secondary origin when the seed carbon and oxygen 
were already present in the star at the time of its formation.
These two channels result in different behaviours of the
abundance of N as a function of metallicity, 
as measured from a primary element such as oxygen.
Primary N production is reflected by a constant N/O ratio 
whereas, when secondary N takes over, the value of N/O is 
expected to rise linearly with O/H.
The existence of a plateau at $\log {\rm (N/O)} \simeq -1.5$
in H~II regions with
$\log {\rm (O/H)} \lsimeq -4.0$ (that is, O/H\,$\lsimeq 1/5$ solar)
is generally interpreted as evidence for
a dominant role of primary production of N at low metallicities 
(e.g. Edmunds \& Pagel 1978; Henry et al. 2000; see also Fig. 3).

There is less agreement, however, as to whether 
this primary nitrogen is produced by massive or intermediate--mass stars. 
The problem has attracted considerable attention in the last few years,
not only because a knowledge of the yields of stars of different masses
is central to chemical evolution models, but also because in principle
in may be possible to use the N/O ratio as a `clock' to date the star
formation activity in galaxies at different cosmological times
(Edmunds \& Pagel 1978; Pettini, Lipman, \& Hunstead 1995; 
Pettini et al. 2002). 
The sensitivity of such a clock depends on the time delay
between the release into the interstellar medium (ISM) of freshly
synthesised N and O. Consider the 
limiting (and no doubt oversimplistic) case of a series 
of bursts of star formation separated
by quiescent periods. Under these circumstances, 
the N/O ratio is at first expected to decrease,
as the O produced by the most massive stars
which explode as supernovae is mixed with the ISM
and the O/H ratio increases accordingly. Only later
will the N/O rise back to the primary plateau,
as the lower mass stars which are the source of primary N
evolve and release the products of their nucleosynthesis.

\begin{figure}[t]
\vskip -0.3cm
\plotfiddle{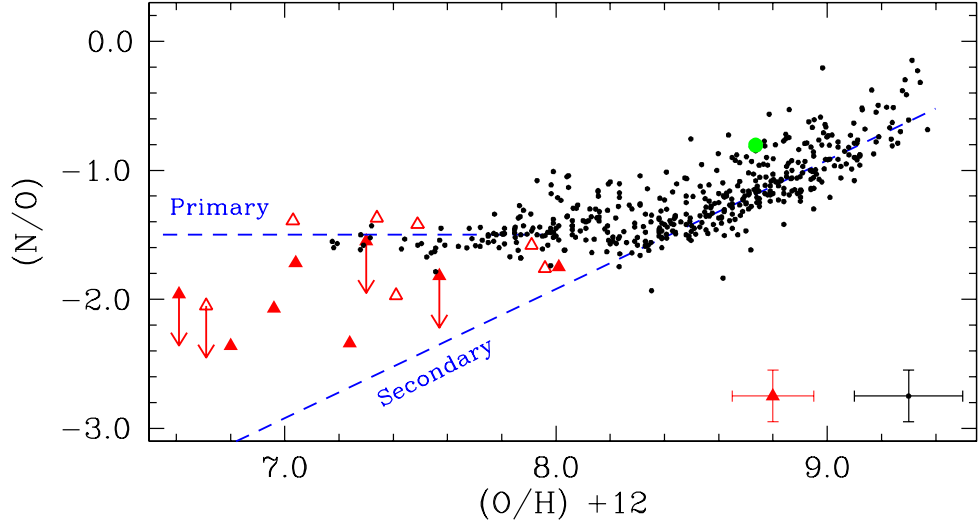}{9cm}{-90}{70}{70}{-290}{350}
\vskip -1.6cm
\caption{Abundances of N and O in extragalactic H II regions (small dots)
and damped Lyman $\alpha$ systems (large triangles). Filled triangles 
denote DLAs where the abundance of O could be measured directly, while open 
triangles are cases where sulfur was used as a proxy for oxygen. 
The large dot corresponds to the solar abundances of N and O from the recent reappraisal by Holweger (2001). 
The dashed lines are approximate representations of the primary and secondary levels of N production. 
The error bars in the bottom right--hand corner give an indication of the typical uncertainties. 
More details regarding this figure and its interpretation may be found in Pettini et al. (2002).}
\end{figure}

Clearly, no such behaviour would result if there is
significant primary N production by massive stars.
Thus, the relative uniformity of the N/O ratios
measured in nearby metal-poor galaxies led Izotov \& Thuan (1999)
to propose that massive stars can perhaps produce significant
amounts of primary N. However, this uniformity could also
be explained if these galaxies have not experienced 
much star formation in the recent past, as indicated by
analyses of their stellar populations.
In contrast, a wide range of values of N/O is
found in star--forming galaxies at intermediate redshifts 
(Contini et al. 2002) and in quasar absorption systems at high 
redshifts (Pettini et al. 2002; Prochaska et al. 2002). 
Even here, though, the interpretation  
is not straightforward. The problem is that
the currently accepted value for the time delay in the 
release of primary N is $\Delta t \simeq 250$\,Myr, 
an interval which seems
too short to explain the high proportion of galaxies
and absorption systems with values of N/O below the primary
plateau (see Fig. 3), as discussed by Pettini et al. (2002).

The value $\Delta t \simeq 250$\,Myr was deduced by 
Henry et al. (2000) from chemical evolution models
which made use of the yields for primary N 
through `Hot Bottom Burning' (see below) in
intermediate mass stars  computed by 
van den Hoek and Groenewegen (1997).
As we now explain, stellar rotation may be the solution---at least in 
part---to this problem, if it shifts the production of 
primary N towards lower mass stars, with longer evolutionary times.
The nucleosynthesis of N during the core He--burning phase of 
metal-poor rotating intermediate-mass stars
has been studied by Meynet \& Maeder (2001, 2002). 
Summarizing their main results:

1) Rotating stellar models which include the same physics which
successfully accounts for: (a) the levels of He and N enrichments measured at the surface
of stars in the Galaxy and in the SMC;
(b) the relative numbers of blue and red supergiants
in the SMC; and (c) the populations of WR stars at solar metallicity, 
naturally lead to the production of primary N
by intermediate mass stars at very low metallicities 
(the metallicity considered in these models was 1/2000 of solar).

2) The process for this production of primary N is different from, 
but not incompatible with, the classical
scenario invoked up to now for intermediate mass stars, 
i.e.  Hot Bottom Burning (HBB).
The mechanism involving rotation 
produces primary N by diffusion of (primary) C and O from
the He--burning core into the hydrogen burning shell. 
This mechanism occurs not only in intermediate mass stars but also, 
although with less efficiency, in massive stars
\footnote{In massive stars, the distance between the He--core and the H--burning shell is 
greater than in intermediate mass stars; moreover the lifetimes are shorter thus, 
for a given initial velocity, the primary N production is much less efficient.}.
The production of primary N by the HBB process occurs in intermediate mass stars only. 
After the core He--burning phase, the star evolves along the asymptotic giant branch. 
There, thermal instabilities drive thermal pulses
which, in some circumstances, may inject C from the He--burning shell
into the base of the convective envelope where the temperature 
may be sufficiently high for the CNO cycle to take place
(see the review on AGB by Frost \& Lattanzio 1995).
The quantities of primary nitrogen produced by this process remain 
difficult to predict (see Marigo 2001; Siess et al. 2002) and
seem to depend heavily on the way mass loss and the third dredge--up mechanism are treated. 
Our models were computed up to the first 3--4 thermal pulses only. 
Since the HBB process is expected to occur at later stages, this means that in our models 
we did not account for any contribution from Hot Bottom Burning.

3) For average rotational velocities on the MS of the order of 230 km s$^{-1}$, it was found
that the most important contributors to the production of primary N are stars in the mass range 
between 3 and 5 M$_\odot$. The massive star contribution is, for this range of velocities, very modest.   

4) The contribution of this process to primary N production is sufficiently efficient to explain
a significant fraction of the primary N observed in low metallicity 
environments (Meynet \& Maeder 2002; Prantzos 2002;
Carigi \& Pettini in this volume).

\begin{figure}[t]  
\plottwo{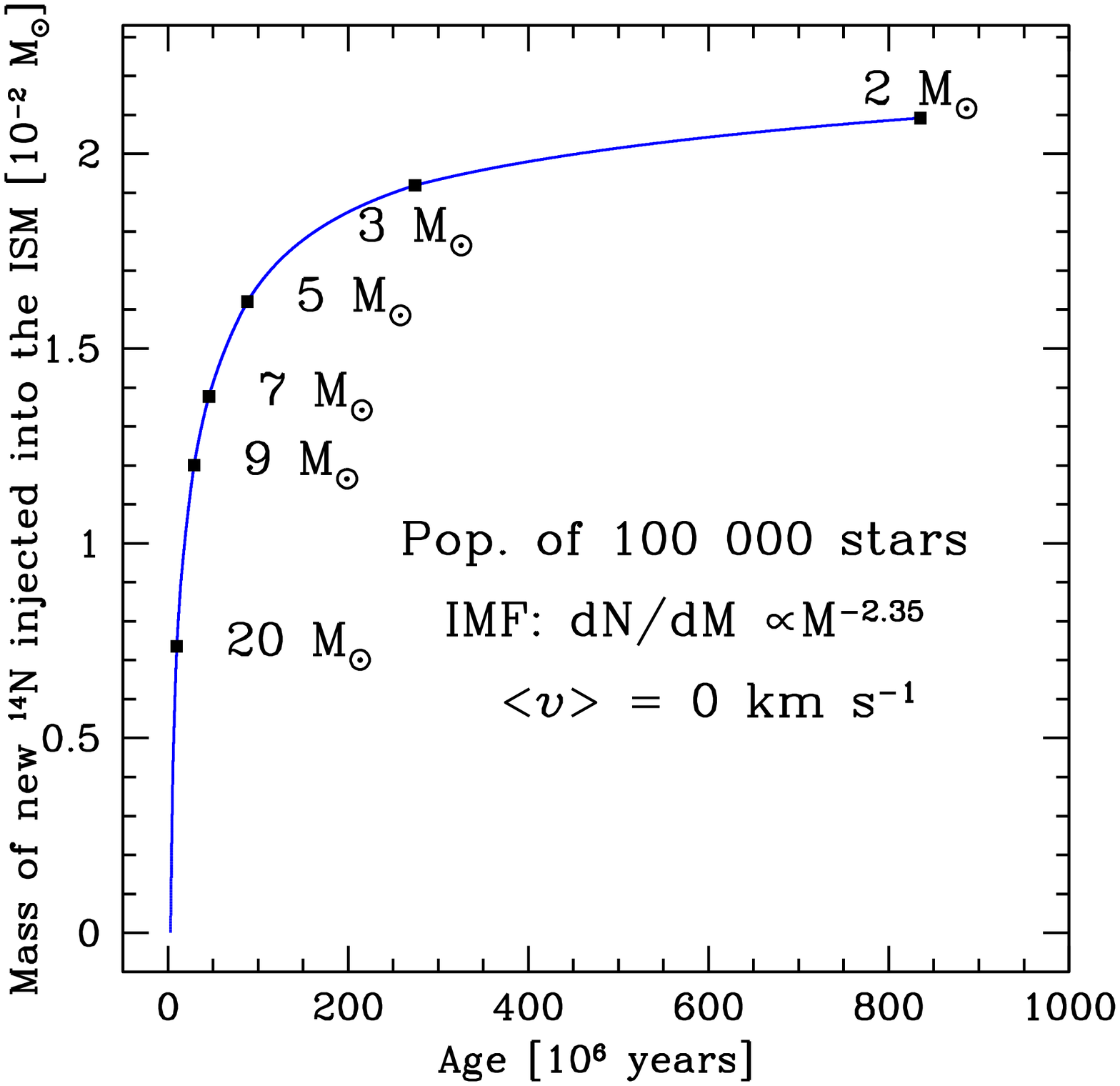}{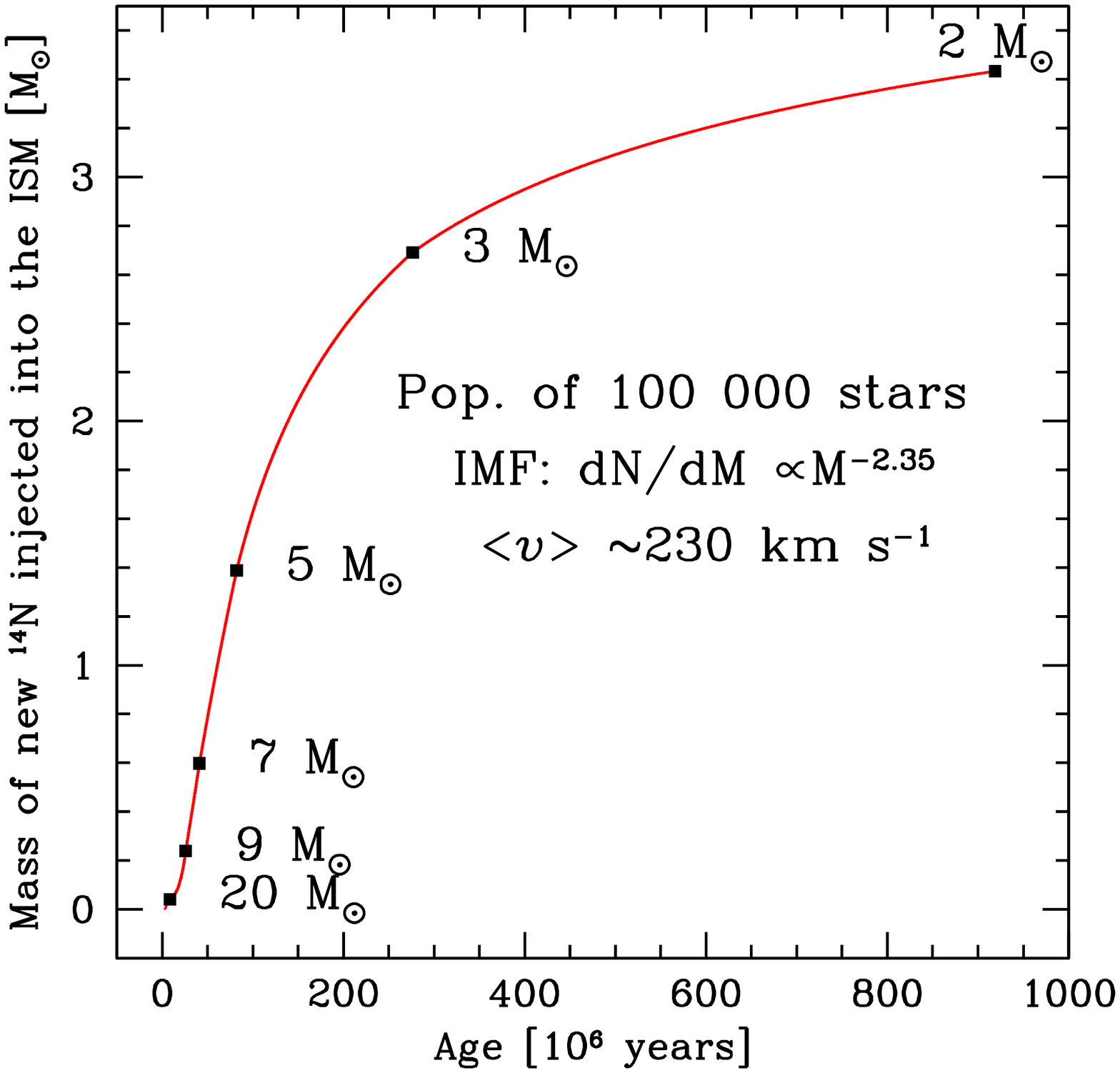}
\caption{Quantities of newly synthesized nitrogen (in solar masses) injected into the interstellar medium by
100 000 stars born at time $t=0$ with initial masses between
0.1 and 60 M$_\odot$. The stellar models have an initial mass fraction of heavy elements equal to 0.00001
(1/2000 the solar metallicity). The left panel shows the situation for non--rotating stellar models, while the right panel
shows the case for rotating stellar models. Note that the vertical axis have different scales in the left and right panel. 
No hot bottom burning process is accounted for. The labels along the curves indicate when a star of a given initial mass contributes.}
\end{figure}

The effects of including stellar rotation in the calculations of the yields of
primary N are illustrated in Fig. 4. Two main differences are noteworthy.
First, the total quantity of primary N injected into the ISM by the rotating models
is two orders of magnitude higher than that obtained with the non--rotating models. 
Second, 95\% of the N is ejected in a little more than 400 Myr by the non--rotating models, 
whereas the time delay increases to 700 Myr when rotation is taken into account.
This larger value of $\Delta t$ is in good agreement with the empirical estimate
derived by Pettini et al. (2002) from consideration of the fraction of damped
Lyman alpha systems (DLAs) exhibiting values of (N/O) below the primary 
plateau at $\log {\rm (N/O)} \simeq -1.5$
in Fig. 3. In the framework of the model described above, these systems are interpreted 
as being observed in the period between the release of O by massive stars 
and of primary N by intermediate mass stars.
By shifting the production of primary N further down the stellar initial mass 
function---and therefore increasing the time delay between the 
O and N enrichment of the ISM following an episode of star formation---stellar 
rotation naturally accounts for: (a) the otherwise
surprisingly high fraction of DLAs with low values of N/O in Fig. 3,
and (b) the finding that most DLAs show little or no enhancement
of the alpha-capture elements relative to Fe (e.g. Vladilo 2002).
Taken together, these two considerations suggest
similar timescales for the production of N and Fe; the standard
view is that approximately two thirds of the latter originates
from Type Ia supernovae.

The agreement between the timescales predicted by the rotating
models and those suggested by the DLA data is likely to be
fortuitous. Observationally, a larger sample of measurements
of N, O and Fe in QSO absorbers is required for trustworthy statistics,
while the models need to take into account a distribution
of values of stellar rotation, as well as Hot Bottom Burning, 
in the calculation of the yields of primary N.
They also need to be extended to a larger set of
metallicities than considered so far.
Nevertheless, the preliminary indications from Fig. 4 are  
that stellar rotation may well provide the `missing clue'
for the interpretation of N abundances.

\acknowledgments{G.M. expresses his deep gratitude to Andr\'e Maeder for the very fruitful collaboration during the past years.}

\end{document}